\documentclass{article}
\usepackage{amsmath,amssymb,latexsym,amsthm,amscd,amsxtra}
\usepackage[all]{xy}
\usepackage{epsfig}
\usepackage{graphicx}

\usepackage{subfigure}

\theoremstyle{definition}

\graphicspath{%
    {converted_graphics/}
    {./}
}

\def\br{\begin{eqnarray}}
\def\er{\end{eqnarray}}
\def\be{\begin{equation}}
\def\ee{\end{equation}}
\begin{document}

\begin{center}
{\Large \bf Barn and Pole paradox: revisited}
\begin{center}
{\large Robert Cacioppo\footnote{e-mail: rcaciopp@truman.edu}$^a$ and Asim Gangopadhyaya\footnote{e-mail: agangop@luc.edu}$^b$}
\end{center}
\begin{tabular}{ll}
a)  &   Division of Mathematics and Computer Science, Truman State University Kirksville, MO 63501\\
b)  &   Department of Physics, Loyola University Chicago, 1032 W. Sheridan Rd., Chicago IL 60660.\\
\end{tabular}
\abstract{We present two different paradoxes related to the length contraction in special relativity and explain their resolution.}
\end{center}

\noindent{\bf Introduction:}

Paradoxes have played great instructive roles in many cultures. They provide an excellent paradigm for teaching concepts that require deep reflection.  Zeno's paradoxes have been discussed for two millennia and they never lose their pedagogical value \cite{Zeno}.
The special theory of relativity is one of the areas of physics where paradoxes are commonly employed for instructive purposes, specially in the context of time dilatation (various incarnations of the well known twin paradox).

For more than one hundred years, notions of time dilation and length contraction have generated tremendous interest among physicists. Despite the presence of numerous books and articles on these subjects \cite{PB1,PB2,PB3,PB4,PB5}, they still retain their mystery and frequently remain misunderstood. Length contraction, originally proposed by George FitzGerald \cite{FitzGerald} and by Hendrik Lorentz\cite{Lorentz}, was not accepted even by leading scientists of the time. The following quote from Joseph Larmor\cite{Brown} shows how people attempted to find alternate explanations for the length contraction as an effect of changes in inter-atomic field strengths: ``... if the internal forces of a material system arise wholly from electromagnetic actions between the system of electrons which constitute the atoms, then the effect of imparting to a steady material system a uniform velocity of translation is to produce a uniform contraction of the system in the direction of motion, of amount $\sqrt{1-v^2/c^2)}$~."

While teaching relativity to freshmen and sophomores, one invariably gets questions such as ``What is the actual length?" The absoluteness of the length and the duration of a time interval is so well ingrained in our psyche that they need to be frequently perturbed by questions that require deeper understanding. We believe that paradoxes that challenge us to rethink about the relativity of spatial intervals and time intervals are exactly the vehicle to bring in glimpses of that enlightenment. Special relativity provides the most effective platform for teaching how to separate real clues from red herrings. It teaches us to see ``reality" as perceived differently by different observers and provides legitimacy for the different measurements of the length (or time interval) as being the ``actual" length or time that each observer measures.

In this paper, we present two thought experiments that at first sight appear different from each other, but are both explained with very similar reasonings. We hope that these two Gedanken experiments and their explanations will help undergraduate students to develop a deeper understanding of the subject and train them to transfer this knowledge to other paradoxes in the area of relativity, or in other branches of learning.

In the first thought experiment, we have a battery that has a length $d$ in its rest frame. The positive and the negative ends of the battery are connected to two metallic hooks that help it hang from a rail. This rail is mostly non-conducting, except for a patch of length $d$ (in the rail's frame), where a piece of copper has been inserted. The battery moves on this rail at high speed. The positive and negative terminals of the battery are always in contact with the rail. However, due to the non-conductivity, the battery does not short-circuit. The question is: What would happen when the battery passes over the conducting part of the rail. According to a linesman in the rail's frame, the battery is shorter than the rail and hence should get short-circuited as it passes over the patch. However, an observer riding with the battery sees the patch to be shorter in length than the battery and hence there is no reason for a short-circuit. Who is correct?

As the second problem, we consider a modified version of the well known ``Pole (or Ladder) and Barn" paradox. We have a fly sitting at the front end of a pole whose rest length is $L$.  The pole is moving towards a barn whose length in its own frame is also $L$. The barn has two sliding doors.  Each door opens as soon as the pole reaches a door and it closes as soon as the rear end of the pole crosses it. Incidentally, this fly suffers from aschluophobia, i.e., fear of darkness. It cannot survive in the dark. However, the fly is not afraid of going through the barn as it knows that the pole, being larger than the barn from the fly's perspective, can never fit inside the barn and the front end of the barn will remain open as long as the fly is inside the barn. As a result, it will not experience total darkness, and hence, will survive the travel through the short tunnel (barn) with opening and closing doors.  However, for a farmer in the barn the pole is much shorter than the barn. He would see the front door close as soon as the the rear end of the pole crosses it.  Since the second door is yet to open, albeit for a very short time, the pole will be completely inside the barn with both doors closed and hence the fly should experience darkness. Thus, the farmer does not expect the fly to survive the ordeal and thinks that it should be dead when the pole emerges from the barn. Who is correct?

\noindent{\bf Analysis of the battery-train:}

\begin{figure}[htb]
\centering
  \includegraphics[width=2.0in]{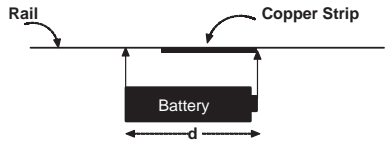}\\
\caption{Moving copper strip as seen from the rest frame of the battery}\label{Battery-perspective}
\end{figure}

Let us consider the battery paradox first. A battery-train is sliding along a non-conducting rail so that its terminal posts
are always in contact with the rail. The following two figures respectively describe the perspectives of a man, at rest with the rail, and the engineer of the battery-train moving with respect to the rail with speed $v$.
\begin{figure}[htb]
\centering
  \includegraphics[width=2.0in]{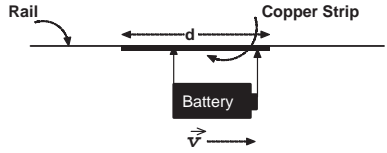}\\
\caption{A moving battery as seen from the rest frame of the rail}\label{Rail-perspective}
\end{figure}
The length of the battery-train in its frame is $d$. There is one section of the rail that is made of copper. The rest length of this
section of rail is also $d$. The chief engineer feels safe including the copper section in the rail, since he finds that when he sets the battery on the rail in motion, the terminals are able to straddle the copper section so that there is no short-circuit. His friend on the ground is still worried - he says that as the battery moves faster along the rail, the distance between the terminals will decrease allowing both to make contact with the copper section, resulting in a short circuit. The chief engineer retorts that in the battery's frame, the copper section will be even shorter so there can be no short circuit as shown in Fig. (\ref{Battery-perspective}). Who is right?

The chief engineer is right! Let us see why there is no short circuit in the rail's frame either. Just to be safe, we will assume that we have a closed loop in a frame if the length of the battery is equal or less than that of the conducting piece. In the rail's frame the speed of the battery is $v$, which is obviously always greater than zero. Let us define $\beta= v/c$, a dimensionless measure for the speed and $\gamma = \frac1{\sqrt{1-\beta^2}}$, a ubiquitous factor of special relativity. The distance between the terminals of the battery in this frame is $d/\gamma$. The amount of time in this frame that both terminals will be in contact with the copper section is
$$\Delta t = \frac1v \left( d- d/\gamma \right)  = \frac{d \, \left(\gamma-1 \right)}{v\gamma}.$$
As shown in Fig. (\ref{Rail-perspective}), the positive terminal contacts the copper section first, followed by the negative terminal. A circuit is not made until the negative terminal reaches the copper section, at which time the information that the negative terminal is now on the copper section must reach the forward (positive) terminal in order for the circuit to be completed. The fastest this information
can travel is $c$ and it must reach the forward terminal within the time $\Delta t$ (that
is, while the forward terminal is still on the copper).

But during this time, the battery will have moved forward a distance $v\, \Delta t$. So, for circuit to be shorted, we must have
$$c\, \Delta t \, \geq \, \frac d\gamma + v\, \Delta t~;$$
i.e., after substituting for the value of $\Delta t $,
$$ \left(c - v\right) \, \frac{d \, \left(\gamma-1 \right)}{v\gamma}\, \geq \, \frac d\gamma ~.$$
This implies that we must have $$\left(1 - \beta  \right)\,\left(\gamma-1 \right) \geq  \beta~.~~$$
Which then implies that
$$\gamma \,\left(1-\beta\right)  = \frac{\sqrt{1-\beta}}{\sqrt{1+\beta}} \geq 1~,$$
which is impossible since we assumed $v>0$ and so $\beta>0$. Hence there will not be a short circuit from the perspective of the railman either.

\noindent{\bf Modified version of the ``Pole (or Ladder) and Barn" paradox:}

\begin{figure}[htb]
\centering
  \includegraphics[width=2.50in]{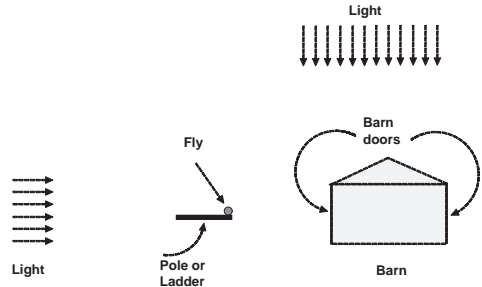}\\
\caption{The ladder and the fly as seen by a farmer in the barn}\label{barn-perspective}
\end{figure}

\begin{figure}[htb]
\centering
  \includegraphics[width=2.50in]{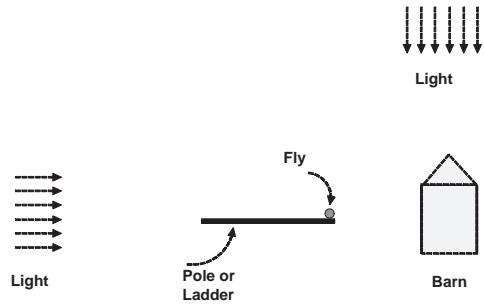}\\
\caption{The barn as seen by the fly}\label{fly-perspective}
\end{figure}

\begin{figure}[htb]
\centering
\subfigure[]{\epsfig{figure=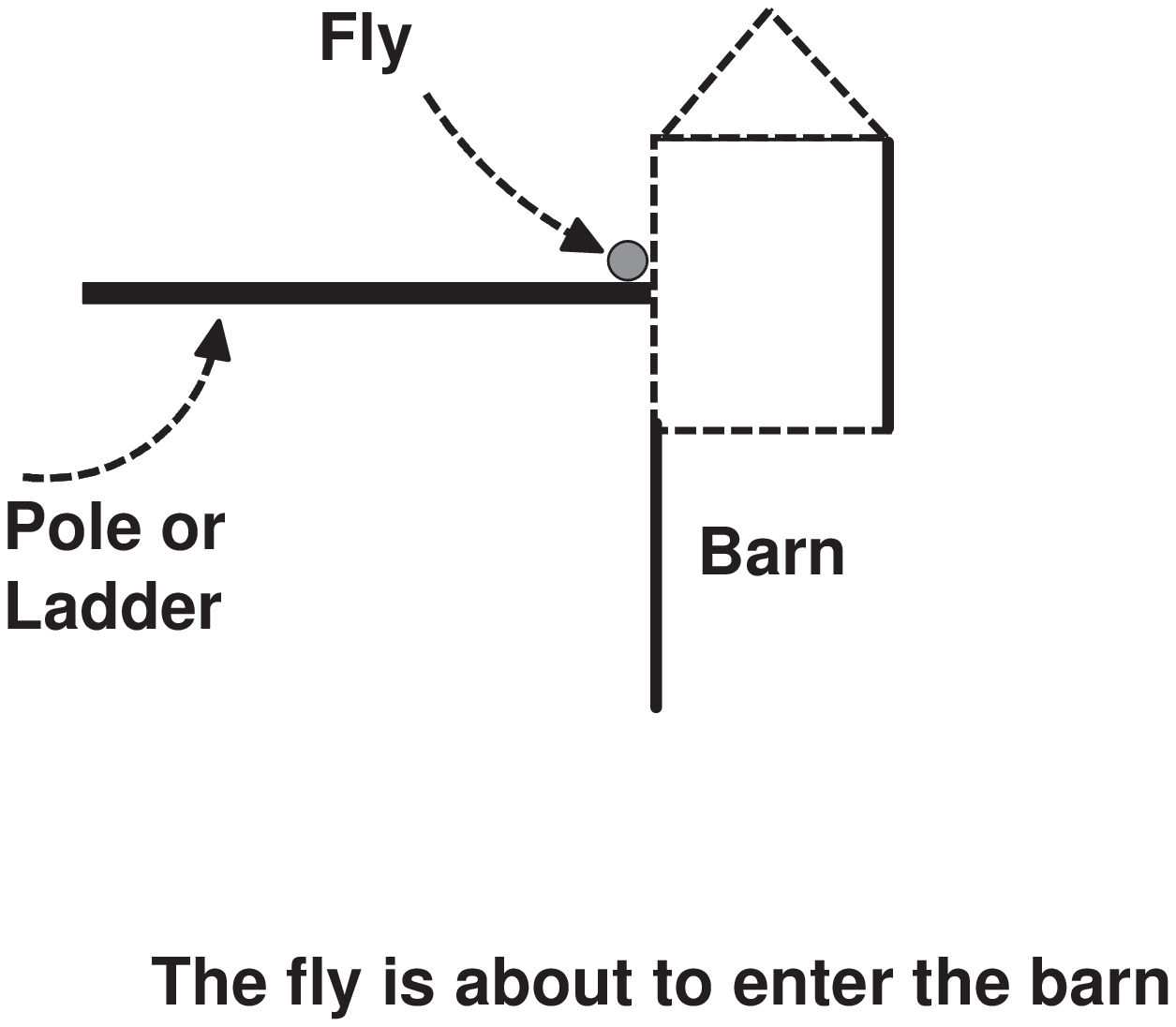,width=.35\textwidth}}
\subfigure[]{\epsfig{figure=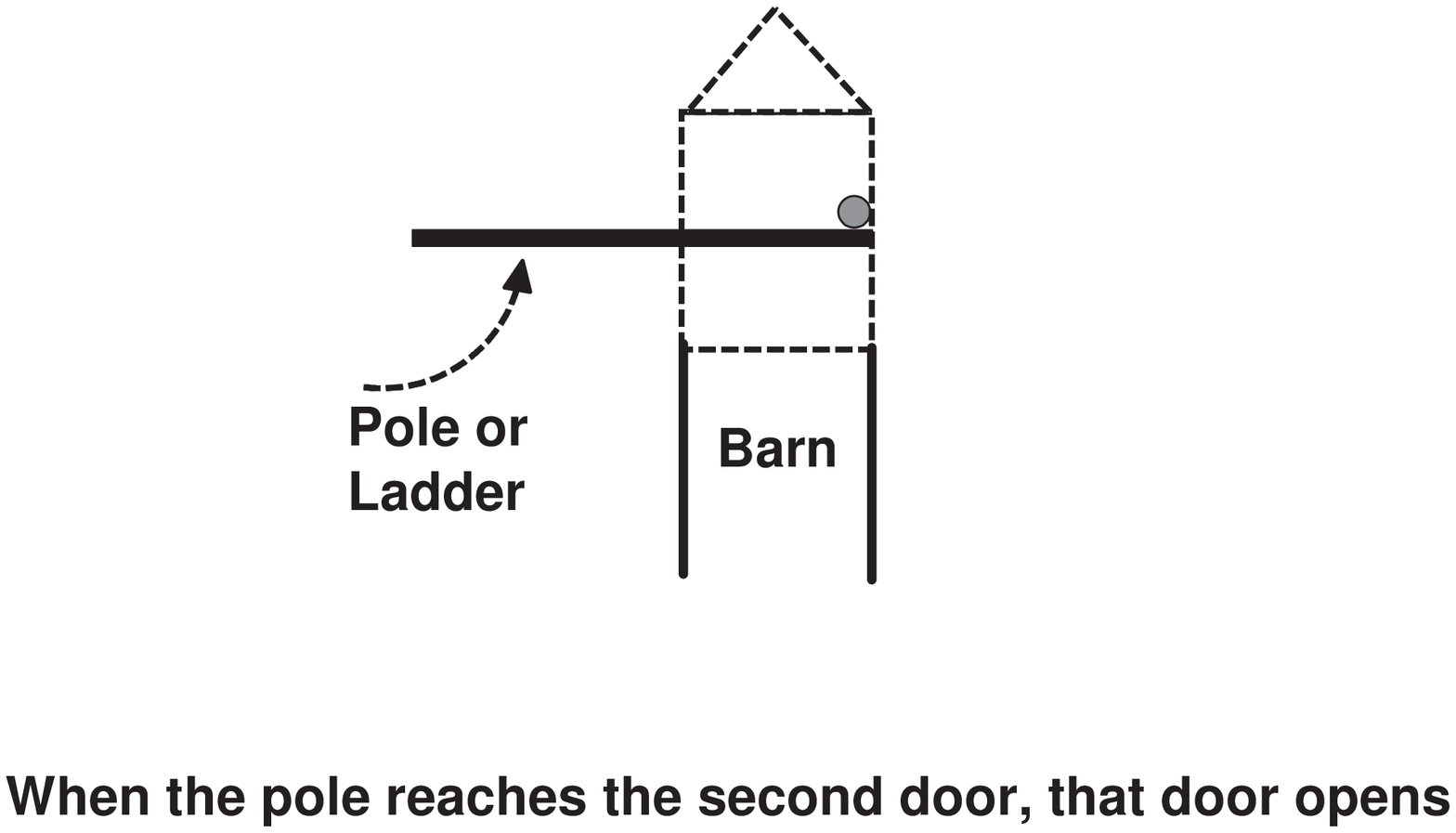,width=.35\textwidth}}\\
{\quad}\\ {\quad}\\
\subfigure[]{\epsfig{figure=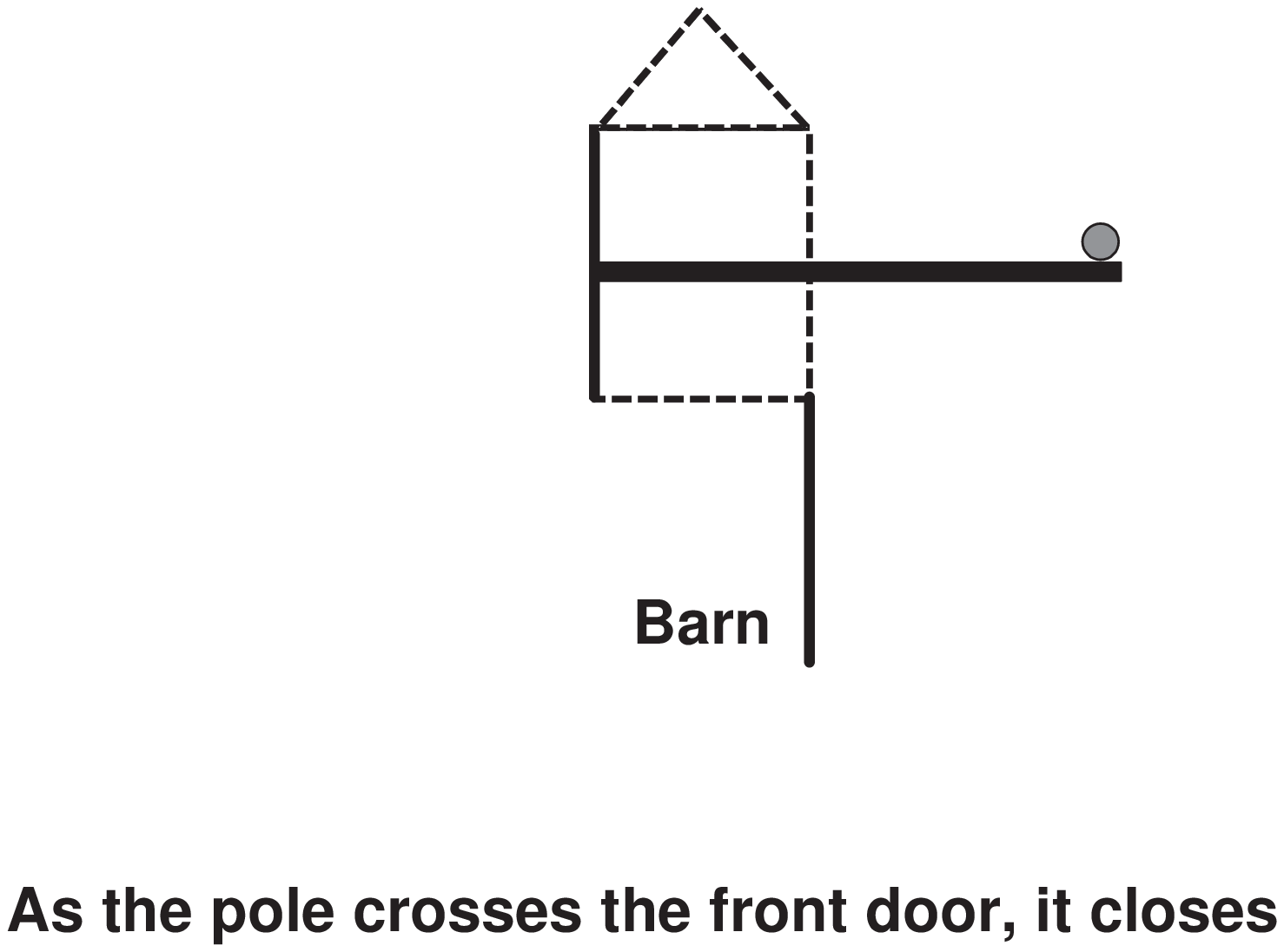,width=.35\textwidth}}
\subfigure[]{\epsfig{figure=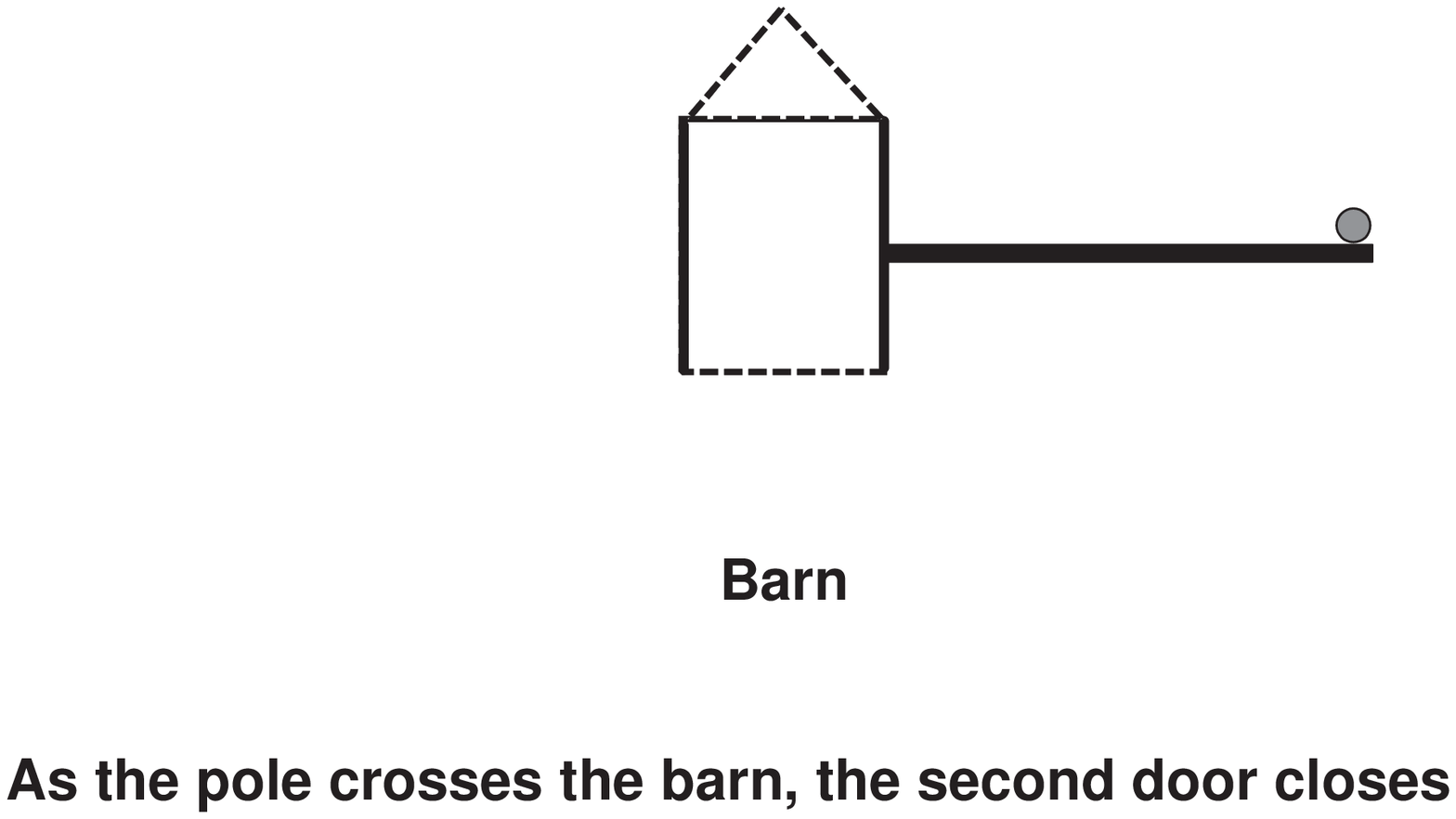,width=.35\textwidth}}
\caption{The sequence of doors opening and closing as seen by the fly}\label{Fly-perspective-of-the-entire-sequence}
\end{figure}

Now, let us visit our second paradox. As the pole approaches the barn, the farmer in the barn sees the pole to have a shorter length than the barn, as shown in Fig. (\ref{barn-perspective}).
We assume that well collimated light is coming vertically down and from the left as shown by the arrows of light.
The perspective of the fly, which finds the barn smaller than the pole, is shown in Fig. (\ref{fly-perspective}).
As mentioned earlier, the fly is sure that it will survive because the pole is never completely inside the barn. The sequence of events seen by the fly is depicted in Fig. (\ref{Fly-perspective-of-the-entire-sequence}).

Since the fly cannot remain alive in one frame and be dead in the other, let us analyze this problem from the perspective of the farmer. As shown in Fig. (\ref{barn-perspective}), the photons are coming from the left, we need to see whether the photon density at the location of fly ever reduces to zero. From the farmer's perspective, the  lengths of the barn and the pole are $L$ and  $L/\gamma$ respectively. After the pole crosses the front door and the door closes, the pole keeps moving inside the barn for the duration of $\overline{\Delta t} = \frac{\left(L-L/\gamma\right)}{v}~.$  During this time, both doors remain closed and the pole is completely inside the barn, and hence, this is the crucial period during which the fly could potentially die unless it can be shown that the photon density seen by the fly would remain non-zero.  It is important to note that we are using darkness as a local concept and not globally defined for all points inside the barn. Photons coming from the left cease to enter the barn as soon as the front door is closed and these photons keep moving away from the front door with speed $c$.

\begin{figure}[htb]
\centering
  \includegraphics[width=2.00in]{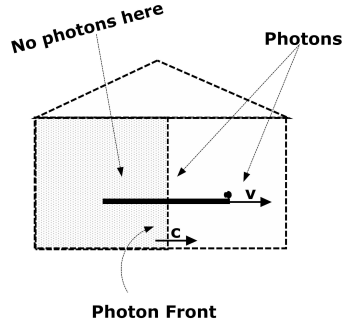}\\
\caption{Receding photon front as seen by a farmer in the barn}\label{barn-perspective2}
\end{figure}

Thus, as shown in Fig. (\ref{barn-perspective2}), a front of photons is created that keeps receding away from the door. If this front, which is moving forward with speed $c$, passes the fly while the fly is still inside the barn, the fly would die.

The time for the front to reach the fly, as seen by the farmer, is $$\Delta t = \frac{\left(L/\gamma\right)}{c-v} $$ and the time for the fly to be out of the barn (after the closing of the front door) is $$\overline{\Delta t} = \frac{\left(L-L/\gamma\right)}{v}~.$$
In the barn's frame, the survival of the  fly is guaranteed if $\Delta t > \overline{\Delta t}$. Thus, both fly and the farmer share the same observation if and only if,
\begin{eqnarray}
\frac{\left(L/\gamma\right)}{c-v} > \frac{\left(L-L/\gamma\right)}{v}~~&; ~~~~~~~~~~~~~~&{\rm i.e.,~if ~and ~only ~if}\nonumber\\
&\nonumber\\
\frac{\left(L/\gamma\right)}{c-v} + \frac{\left(L/\gamma\right)}{v}> \frac{L}{v}~~&; ~~~~~~~~~~~~~~&{\rm i.e.,~if ~and ~only ~if}\nonumber\\
&\nonumber\\
\frac{1}{c-v} + \frac{1}{v}> \frac{\gamma}{v}~~&; ~~~~~~~~~~~~~~&{\rm i.e.,~if ~and ~only ~if}\nonumber\\
&\nonumber\\
\frac{c}{c-v} > \frac1{\sqrt{1-\beta^2}}~~&; ~~~~~~~~~~~~~~&{\rm i.e.,~if ~and ~only ~if}\nonumber\\
&\nonumber\\
\frac{1}{1-\beta} > \frac1{\sqrt{(1-\beta)(1+\beta)}}~~&; ~~~~~~~~~~~~~~&{\rm i.e.,~if ~and ~only ~if}\nonumber\\
&\nonumber\\
\frac{\sqrt{(1+\beta)}}{\sqrt{(1-\beta)}}>1 ~~&; ~~~~~~~~~~~~~~&{\rm i.e.,~if ~and ~only ~if}~\beta>0~.\nonumber\end{eqnarray}
But, $\beta$ is just the ratio of the speed of the pole and the speed of light and is always restricted to the domain $0\le \beta<1$.  Thus, for all non-zero values of the speed $v$, the fly will survive the travel through the barn.

\noindent{\bf Conclusion:}
Thus, we presented two perplexing examples of length contraction. Both examples are built by adding additional structures to the well known Pole and Barn paradox.  Explanation of both paradoxes requires a very similar reasoning, which is not unexpected, as both examples can be shown to be mathematically isomorphic to each other. One of the main features of these examples is that their resolution requires that the speed of propagation of all interactions be limited by the speed of light, a major tenet of the special theory of relativity. We feel that these types of paradoxical problems and their explanations provide excellent paradigms for instruction in special relativity.

\noindent{\bf Acknowledgement:}
We would like to thank the referee for several constructive comments, and Prof. Thomas T. Ruubel for a very careful reading of the manuscript and helpful suggestions.

\end{document}